\documentclass[aps,pre,preprint,superscriptaddress,showpacs,showkeys]{revtex4-1}

\usepackage{amsmath, amsfonts, amsthm, amssymb, mathrsfs, enumerate, graphicx, cases, hyperref}
\usepackage[T1]{fontenc}

\newcommand{\R}{{\mathbb R}}

\begin{document}


\title{Asymptotically simple spacetimes and mass loss due to gravitational waves}


\author{Vee-Liem Saw}
\email[]{VeeLiem@maths.otago.ac.nz}
\affiliation{Department of Mathematics and Statistics, University of Otago, Dunedin 9016, New Zealand}


\date{\today}

\begin{abstract}
The cosmological constant $\Lambda$ used to be a freedom in Einstein's theory of general relativity, where one had a proclivity to set it to zero purely for convenience. The signs of $\Lambda$ or $\Lambda$ being zero would describe universes with different properties. For instance, the conformal structure of spacetime directly depends on $\Lambda$: null infinity $\mathcal{I}$ is a spacelike, null, or timelike hypersurface, if $\Lambda>0$, $\Lambda=0$, or $\Lambda<0$, respectively. Recent observations of distant supernovae have taught us that our universe expands at an accelerated rate, and this can be accounted for by choosing $\Lambda>0$ in Einstein's theory of general relativity. A quantity that depends on the conformal structure of spacetime, especially on the nature of $\mathcal{I}$, is the Bondi mass which in turn dictates the mass loss of an isolated gravitating system due to energy carried away by gravitational waves. This problem of extending the Bondi mass to a universe with $\Lambda>0$ has spawned intense research activity over the past several years. Some aspects include a closer inspection on the conformal properties, working with linearisation, attempts using a Hamiltonian formulation based on ``linearised'' asymptotic symmetries, as well as obtaining the general asymptotic solutions of de Sitter-like spacetimes. We consolidate on the progress thus far from the various approaches that have been undertaken, as well as discuss the current open problems and possible directions in this area.\end{abstract}

\keywords{Gravitational waves, mass-loss formula, Bondi-Sachs mass, cosmological constant, de Sitter, null infinity}

\maketitle


\section{Introduction}\label{Section1}

Albert Einstein's 1915 theory of general relativity (GR) has replaced the centuries-old Newtonian theory of gravitation, backed by numerous affirmative experimental results. The present time is an exciting period to be working in this field, especially when we have just celebrated the first ever direct detection of gravitational waves \cite{LIGO,LIGO2}. Being significantly more complicated, GR's field equations give rise to many bizarre new mathematical solutions not contained in Newton's simple differential equation. The latter treated gravitation as a \emph{force}, relating the gravity of a massive object (law of universal gravitation) to the dynamical behaviour of another massive object (second law of motion) \footnote{The term ``massive'' is used in this review article to mean that the object being referred to is not massless.}.

With a disparate view of gravity being manifested as the curvature of a \emph{spacetime pseudo-Riemannian manifold}, Einstein's theory has led to novel and unprecedented theoretical phenomena. For instance, spacetime manifolds can have peculiar structures with non-trivial topologies allowing for the possibility of a so-called ``shortcut through spacetimes''. Such a construction of \emph{traversable wormholes} was shown to be theoretically plausible by Morris and Thorne \cite{Kip}, if one allows for the existence of ``exotic matter'' having unconventional physical properties which violate the energy conditions. Interestingly, there are designs of such curved traversable wormholes with safe geodesics through them such that a traveller need not directly encounter these exotic matter \cite{Vee2012,Vee2013}. Apart from spatial shortcuts, GR also permits time travel as some spacetimes have been found to contain \emph{closed timelike curves} \cite{CTC,Vee2015,VUW} \footnote{This novel method of constructing spacetimes by generating manifolds of revolution around a given curve was inspired by the development of helicalised fractals \cite{Vee2013b}, where a curve is replaced by another curve that winds around it. Essentially, 2-surfaces are strung together along a given smooth curve to produce a 3-manifold, thereby providing a means of foliating such 3-manifolds. Applications of this method so far, have been to construct curved traversable wormholes in Refs. \cite{Vee2012,Vee2013} as well as to solve the linearised vacuum GR equations in Ref. \cite{Vee2015}.}. In fact, an early solution to GR that sparked questions on its physical meaning as well as perhaps cast serious doubts on whether GR is a viable theory of gravity is the Schwarzschild solution representing a spherically symmetric \emph{black hole}, where the spacetime manifold is \emph{non-regular} at one point \cite{Gra}. Another early solution was derived by Einstein himself, after solving his field equations to first order and arriving at \emph{wavelike solutions} \cite{Einquad}.

Whilst the above four phenomena --- wormholes, closed timelike curves, black holes, gravitational waves, are theoretically predicted by Einstein's theory of general relativity, the prime question of interest to physicists is certainly whether these represent \emph{actual physical reality of our universe} or are merely mathematical curiosities. Shortcuts through spacetime have been a favourite key plot for science fiction \footnote{In fact, Morris and Thorne's work on traversable wormholes \cite{Kip} was motivated by Carl Sagan's enquiry on whether our best known theory (circa 1985) allowed for such possibilities, as he was writing his now famous book \emph{Contact}.}. However, the mandatory presence of even an arbitrarily infinitesimal amount of exotic matter to support such traversable wormholes \cite{Vis2} would arguably veto its physical relevance, in spite of the Casimir effect being a known example of such exotic matter \cite{Exo1,Exo2,Exo3}. Also, the notion of time travel introduces awkward paradoxes and one may wish to expostulate with solutions containing closed timelike curves as being \emph{un}physical. Nevertheless, black holes have been accepted as real astrophysical objects --- in particular, the first direct detection of gravitational waves from a compact binary system shows that this source must be highly compact objects consisting of a pair of black holes \cite{LIGO}.

The question on whether gravitational radiation is a genuine physical phenomenon or just a coordinate/gauge artefact took nearly 50 years to resolve. Einstein's discovery from the linearised field equations meant that those represent weak gravitational fields and it may well be possible that higher order contributions of the coupled set of non-linear partial differential equations would cancel out such wavelike effects. Incidentally, one major difficulty in describing the energy carried by gravitational waves has to do with the key physical underpinning of GR itself, viz. the \emph{equivalence principle}, which allows for gravitational effects to be \emph{locally eliminated} \cite{Hobson}.

It was finally accepted by most physicist in the 1960-ies that gravitational waves constitute a real physical entity, when Bondi et al. showed that the total mass-energy of an isolated gravitating system must decrease when it emits gravitational radiation, i.e. gravitational waves carry energy away from the massive source \cite{Bondi60,Bondi62}. By starting off with an ansatz for an axisymmetric spacetime involving outgoing null cones defined by constant values of the coordinate $u$, an asymptotic expansion towards large distances $r$ away from the isolated system was carried out. The Bondi mass-loss formula was then obtained from one of the ``supplementary conditions'' (arising from the Bianchi identities) whose $r^{-2}$ factor cancelled out \cite{Bondi62}. Very shortly after, Sachs dealt with the general asymptotically flat spacetimes without axisymmetry \cite{Sachs62}.

Incidentally, this Bondi-Sachs mass can be defined from a Hamiltonian framework \cite{Poisson,BLY}, and this shows that it is the total mass-energy within a shell of radius $r$, with $r$ taken to infinity along a null direction --- i.e. at \emph{null infinity}, $\mathcal{I}$. This circumnavigates the issue of \emph{locally} eliminating the effects of gravity due to the equivalence principle, since the Bondi-Sachs mass concerns with the total mass-energy of the \emph{entire} (asymptotically flat) spacetime (i.e. it is generally not possible to eliminate the effects of gravity at more than one single point) \footnote{If this shell is taken to \emph{spatial infinity} instead, then one arrives at the ADM mass, which remains constant even though a system loses energy due to gravitational radiation \cite{adm,ADMDDD,200years,Poisson}.}.


Around that same period (1960-ies), Newman and Penrose (NP) came up with an equivalent formulation of GR expressed in terms of 38 NP equations \cite{newpen62}. Newman and Unti then solved these equations for asymptotically flat spacetimes, using a similar setup involving outgoing null cones given by constant values of the coordinate $u$ \cite{newunti62}. With this, the Bondi mass-loss formula can be expressed as
\begin{eqnarray}\label{Bondimasslossflat}
\frac{dM_B}{du}=-\frac{1}{A}\oint{|\dot{\sigma}^o|^2d^2S},
\end{eqnarray}
where $M_B$ is the Bondi mass, $A=4\pi$ (the area of the unit sphere), and $\sigma^o$ is the leading order term of the complex spin coefficient $\sigma$ when expanded as inverse powers of $r$. The dot denotes derivative with respect to $u$, and a non-zero $\dot{\sigma}^o$ is interpreted as gravitational radiation being emitted by the isolated system. The integral is carried out over a compact 2-surface of constant $u$ on $\mathcal{I}$.

Details on some related history may be found in Ref. \cite{Fra04}, a description of the Bondi-Sachs formalism may be found in Ref. \cite{MadWini}, with that on the NP formalism available in Ref. \cite{NP}.

\subsection{Current research: Bondi mass and mass loss due to gravitational waves with $\Lambda>0$}\label{Section1A}

Now, we observed that our universe expands at an accelerated rate \cite{cosmo1,cosmo2}, and a simple way of taking this into account would be to have a positive cosmological constant $\Lambda>0$ in the Einstein field equations. The presence of $\Lambda>0$ however, would alter the conformal structure of spacetime where null infinity is a \emph{spacelike hypersurface}, instead of a null hypersurface when $\Lambda=0$ \cite{Pen65}. A great deal of work to describe the energy carried by gravitational waves with $\Lambda>0$ has been ongoing over recent years, invoking a raft of different methods \cite{Vee2016,Vee2017,Vee2017b,Vee2017c,Szabados,Chrusciel,chi1,chi2,ash1,ash2,ash3,ash4,gracos1,gracos2,Zhang}. One way of going about this is to take the NP equations with a cosmological constant $\Lambda$ and solve it asymptotically in a manner similar to how Newman and Unti did it for asymptotically flat spacetimes. This behaviour of asymptotically empty (anti-)de Sitter spacetimes was worked out by Saw in Ref. \cite{Vee2016}, with those involving Maxwell fields given in Ref. \cite{Vee2017}. Hitherto, this appears to be the only proposal of a Bondi mass \emph{together with} a mass-loss formula (therefore, a deduction of the energy carried by gravitational waves) that has a cosmological constant --- within the \emph{full non-linear Einstein's theory of general relativity}, as well as having the appealing form which directly reduces and clearly relates to the well-known asymptotically flat case when $\Lambda$ is set to zero (e.g. giving Eq. (\ref{Bondimasslossflat}) above). In addition, Saw also obtained the peeling properties of the Weyl and Maxwell spinors with $\Lambda$, just like how Newman and Penrose derived it rather easily with the NP formalism for the $\Lambda=0$ version \cite{newpen62}. This provides an explicit derivation of the peeling property based on the physical spacetime, complementing the same result obtained by Penrose for asymptotically simple spacetimes (i.e. with a cosmological constant) using the conformal approach \cite{Pen65}. Apart from Saw's work, there are nevertheless, two other proposals for the Bondi mass with $\Lambda$ given by Szabados and Tod \cite{Szabados}, as well as by Chru\'{s}ciel and Ifsits \cite{Chrusciel}, using the full non-linear theory.

These three approaches by Saw, Szabados-Tod, Chru\'{s}ciel-Ifsits, are independent and seem to be unrelated: 1) Saw solved the NP equations with $\Lambda$ asymptotically \`{a} la Newman and Unti purely in physical spacetime (i.e. without directly invoking the conformal structure); 2) Szabados and Tod used twistor methods; whilst 3) Chru\'{s}ciel and Ifsits essentially also solved the Einstein equations asymptotically (and so would in principle be equivalent to what Saw produced). However, instead of solving the NP equations, they made use of the Fefferman-Graham expansions that assume a smooth conformal compactifiability of vacuum spacetimes \cite{FG1,FG2}. Furthermore, their key result is referred to as a ``balance formula'' which gives an expression for the mass, instead of a mass-loss equation. In other words, their ``balance formula'' as presented in Eqs. (5.56) or (9.1) in Ref. \cite{Chrusciel} is not of the ``mass-loss'' type in Eq. (\ref{Bondimasslossflat}) above. Anyway, it should be noted that the expression for the mass that they obtained is \emph{manifestly positive-definite}.

At present, it is not obvious how to draw meaningful comparisons between these different proposals which are based on completely distinct ideas. In particular, they all use disparate notations and techniques, which obfuscate how a set of terms from one work would relate to the others. Nevertheless, there is one worthy observation: In Saw's proposal for the Bondi mass with $\Lambda>0$ presented in Eq. (126) in Ref. \cite{Vee2016} (or see Eq. (\ref{topsph}) in Section \ref{Section3} below), there are two new terms involving $\Lambda$. One of these terms comprises a surface integral over a topological 2-sphere of constant $u$ on $\mathcal{I}$ followed by another integral over $u$. Coincidentally, the ``balance equation'' for the mass by Chru\'{s}ciel and Ifsits has a term which is interpreted as a \emph{renormalised volume of the null hypersurface}. This is certainly one point that should be more closely investigated. Note also that whilst the Chru\'{s}ciel-Ifsits mass is manifestly positive-definite, the Saw mass is proposed to be of such a form in Eq. (\ref{topsph}) so that the \emph{mass-loss formula with $\Lambda>0$ is manifestly positive-definite} (in the absence of incoming radiation), i.e. Eq. (\ref{Bondimassloss}) below. Ergo, this guarantees that the \emph{energy carried by gravitational waves is strictly non-negative}.

In the asymptotically flat case incidentally, the original approach to this problem by Bondi et al. \cite{Bondi62} also carried out an asymptotic expansion towards null infinity $\mathcal{I}$ for an axisymmetric system (with Sachs providing the analysis for general asymptotically flat spacetimes \cite{Sachs62}). The Bondi-Sachs way is equivalent to the Newman-Penrose-Unti method \footnote{There is an anecdote given in Ref. \cite{MadWini}, describing the independent and parallel developments by Bondi et al., versus Newman, Penrose, Unti, back in the 1960-ies. Bondi would ask at meetings: ``Are you a qualified translator?'', as these two approaches used notations and formalisms which were completely different.}. As it turns out, He and Cao have strived to mimic the Bondi approach \cite{chi1}, and further succeeded to extend their study with Jing to include Maxwell fields \cite{chi2}. They proposed a new leading term in the Bondi ansatz for the axisymmetric metric to account for $\Lambda$, and calculated the dyad component of the Weyl spinor $\Psi_4$ to show that it has an $r^{-1}$ leading term containing the usual Bondi news plus new terms due to $\Lambda$. The appearance of the Bondi news signifies the usual notion of gravitational radiation carrying energy away from the source, with corrections due to the cosmological constant.

Apart from that, they are perhaps the first to calculate the metric for $\mathcal{I}$ with $\Lambda$ (in the axisymmetric case), which evidently showed that it is \emph{non-conformally flat} when the Bondi news is non-zero --- i.e. the Cotton-York tensor for $\mathcal{I}$ vanishes if and only if the Bondi news is zero. The identical axisymmetric metric for $\mathcal{I}$ is also independently produced by Saw as it follows from 2 of the 38 NP equations (the metric equations for $D'\xi^\mu$), and this similarly holds for the general non-axisymmetric case \cite{Vee2016}. This is because one cannot specify all the compact 2-surfaces of constant $u$ on $\mathcal{I}$ to be round 2-spheres (see Sections \ref{Section2C} and \ref{Section2D} below). The paper by He and Cao \cite{chi1} however, did not seem to go on and use the supplementary conditions, i.e. the Bianchi identities, to get the mass-loss formula. If one would actually do that, the expected result would be the mass-loss formula obtained by Saw, since it is the same Bianchi identity that gives rise to a derivative of the ``mass aspect'' with respect to the retarded null coordinate $u$ --- regardless of whether it uses the Bondi-Sachs or the Newman-Penrose formalism (see Eq. (\ref{masslosspreintegration}) below in Section \ref{Section3}). Incidentally, both Saw \cite{Vee2017} and He et al. \cite{chi2} independently found that the Maxwell fields do not affect the structure of $\mathcal{I}$ --- i.e. electromagnetic radiation carrying energy away from an isolated source does not lead to $\mathcal{I}$ being non-conformally flat.

Whilst the work by Saw and He et al. are based exclusively on the physical spacetime, Penrose had illustrated the power of conformal compactification back in the 1960-ies \cite{heledi,pen11}. This is undoubtedly one prime avenue for researchers to study the problem of formulating the Bondi mass and the mass-loss formula with $\Lambda>0$.

\subsection{Conformal structure}\label{Section1B}

Prior to the discovery in the 1990-ies supporting a need for $\Lambda>0$, there have been some investigations of the mass with a cosmological constant \cite{AD,AD2}, though not of a Bondi-Sachs type that would decrease when an isolated system radiates gravitational waves. A first clear desire to obtain the Bondi mass for a universe with $\Lambda>0$ might have been expressed in an article by Penrose \cite{Pen2011}. His motivation arose from his proposal on the ``conformal cyclic cosmology'' (CCC) \cite{Pen2010}, which relies heavily on the conformally compactified asymptotically de Sitter spacetimes having a spacelike $\mathcal{I}$. According to his CCC proposal, the future null infinity of a previous aeon would form the big bang for the succeeding aeon, so gravitational radiation from the former universe would propagate into the next one. Whilst this would intriguingly lead to some testable predictions, it also necessitates the formulation of a well-defined notion of a Bondi-type mass for asymptotically de Sitter spacetimes in order to correctly describe the energy carried by gravitational waves in a universe with $\Lambda>0$. Penrose's 2011 paper became somewhat of a precursor to the work by Szabados and Tod \cite{Szabados} that evolved into using twistor methods. Incidentally, the latter demonstrated that Penrose's suggestion for a definition of a cosmological mass based on a charge integral of curvature \cite{Pen2011} does not have the ``rigidity property'' (so it could be zero even for non-trivial setups).

In their extensive and technical paper \cite{Szabados}, Szabados and Tod began with examining the conformal properties of asymptotically de Sitter spacetimes and reported that with $\Lambda>0$, it is generally not possible to foliate the spacelike $\mathcal{I}$ by round 2-spheres. This has the ramification that $\mathcal{I}$ would be non-conformally flat, unless one forces all these compact 2-surfaces to be round 2-spheres (which is shown by Saw in Ref. \cite{Vee2016}, based on the physical spacetime --- see also Section \ref{Section2C} below). Ostensibly surprising initially, Ashtekar, Bonga and Kesavan who also worked with the conformal structure \cite{ash1} discovered that imposing conformal flatness on $\mathcal{I}$ would imply that the energy carried by gravitational waves away from the isolated source is \emph{zero}, as it corresponds to having the magnetic part of the leading order asymptotic Weyl curvature to be zero \footnote{This is analogous to the study on asymptotically anti-de Sitter spacetimes back in the 1980-ies by Ashtekar and Magnon \cite{ashmag}, where conformal flatness of the timelike $\mathcal{I}$ was assumed. This allowed for conserved quantities to be written as integrals involving asymptotic Killing fields of the anti-de Sitter background. These quantities are however, \emph{absolutely conserved} (in the absence of matter fields near $\mathcal{I}$), i.e. there is \emph{no Bondi news} analogous to the asymptotically flat case. Well, the asymptotic solutions by Saw \cite{Vee2016,Vee2017} show that conformal flatness of $\mathcal{I}$ for any non-zero $\Lambda\in\R$ \emph{cannot} be imposed by hand, especially if one wishes to describe gravitational waves. This is because \emph{the radiating isolated source would determine the structure of $\mathcal{I}$ when $\Lambda\neq0$}.}.

Also from the conformal approach in studying de Sitter-like spacetimes, Szabados and Tod derived the fall-offs for the metric and spin coefficients of asymptotically de Sitter spacetimes. This proved to be highly beneficial, as the ansatz for these fall-offs made by Saw purely by investigating the Schwarzschild-de Sitter spacetime (only the physical one, but not the conformal one) \cite{Vee2016} turned out to be consistent with those given by Szabados and Tod --- providing validation that such a stipulation are merely fixing gauge freedoms and the corresponding asymptotic solutions are general. The eventual key result of Szabados and Tod was to obtain a Bondi-type mass for de Sitter-like spacetimes using twistor methods and the Nester-Witten 2-form \cite{Pen88} \footnote{There is an integral formula on a hypersurface that was derived by Frauendiener based on the Nester-Witten identity \cite{Fra97}, and is expressed using the NP spin coefficients. One of the several applications of this integral formula on a hypersurface is to produce the Bondi mass-loss formula for asymptotically flat spacetimes, i.e. Eq. (\ref{Bondimasslossflat}), by adapting this into a null hypersurface that approaches $\mathcal{I}$ and plugging in the Newman-Unti asymptotic solutions for asymptotically flat spacetimes \cite{newunti62}. From this perspective, the mass-loss formula may be thought of as a geometric property --- arising from this Nester-Witten identity. With the cosmological constant present, one can adapt the spacelike (or timelike, depending on the sign of $\Lambda$) hypersurface to approach $\mathcal{I}$, and apply the asymptotic solutions found by Saw \cite{Vee2016,Vee2017}. This will be discussed elsewhere.}. Furthermore, they succeeded in proving that this expression for the mass is positive-definite and has the desired \emph{rigidity} property (so it does not just vanish for some non-trivial systems).

\subsection{Linearised theory}\label{Section1C}

After showing that a conformally flat $\mathcal{I}$ for $\Lambda>0$ does not allow energy to be carried away by gravitational waves \cite{ash1}, Ashtekar, Bonga and Kesavan subsequently worked this out explicitly in the linearised version, as well as in the Maxwell theory when the magnetic field is set to zero at $\mathcal{I}$ \cite{ash2}. To do so, they employed the covariant phase space formalism \cite{200years}, making use of the Killing fields of the de Sitter background. As a natural next step, these authors went ahead to produce the quadrupole formula for linear fields \cite{ash3}, thereby generalising Einstein's result \cite{Einquad} to now include a cosmological constant.

Notable results from their work are: 1) Gravitational and Maxwell radiations may purportedly carry an arbitrarily negative energy away from the isolated system, though for physically relevant systems the energy carried away is necessarily positive-definite. They argued that the ``negative energy'' is actually incoming radiation from elsewhere, associated with the timelike Killing vector field becoming spacelike and past-directed beyond the isolated system's cosmological horizon. In Saw's full mass-loss formula \cite{Vee2016} (see Eq. (\ref{0BondimasslossGraonly}) below for purely gravitational waves, i.e. without Maxwell fields), there is an explicit term containing the dyad component of the Weyl spinor $\Psi^o_0$ (where the superscript $^o$ denotes its leading order term), clearly indicating that a non-positive-definite contribution is due to \emph{incoming radiation} from elsewhere, thereby affirming this deduction \footnote{Note that $\Psi_4$ and $\Psi_0$ represent outgoing and incoming gravitational waves, respectively. Similarly, $\phi_2$ and $\phi_0$ represent outgoing and incoming electromagnetic waves, respectively.}. Furthermore, this is particularly perspicuous for the electromagnetic (EM) case as can be seen in Eq. (\ref{0BondimasslossMaxonly}) below \cite{Vee2017}, where outgoing EM radiation carries positive-definite energy away from the isolated source and incoming EM radiation from elsewhere carries positive-definite energy to the source. 2) Besides that, Ashtekar et al.'s linearised quadrupole formula showed that the corrections due to the cosmological constant are in powers of $\sqrt{\Lambda}$, and it plays a negligible role such that LIGO would not be able to detect any difference between $\Lambda>0$ and $\Lambda=0$. A summary of their results may be found in Ref. \cite{ash4}.

Apart from papers by Ashtekar, Bonga and Kesavan \cite{ash2,ash3}, there are also others who reported results based on the linearised gravitational theory \cite{gracos1,gracos2}. For instance, Bishop worked within the Bondi-Sachs framework (having the same $u,r,x^\mu$ coordinates used by Saw \cite{Vee2016}, i.e. the metric for (anti-)de Sitter spacetime is Eq. (\ref{puremet}) below in these coordinates) to construct exact solutions to the linear theory and got an expression for the energy of the radiation measured by distant observers \cite{gracos1}. The corrections due to the cosmological constant are shown to be of positive integer powers of $\Lambda$. This is consistent, as Saw showed from the full non-linear theory, that an expansion for tiny $\Lambda$ would also give corrections to the mass-loss formula in positive integer powers of $\Lambda$. On top of that, Bishop arrived at the expected peeling property of the Weyl spinor, and that the corrections due to $\Lambda$ are ignorable for systems of interest (which LIGO would be able to look out for).

In contrast to the full theory dealt with by Saw, the linearised theory by Bishop does not reflect the nature of $\mathcal{I}$ being non-conformally flat and the perturbations are carried out with respect to the de Sitter background. Perhaps in a future study, one may consider a linearised theory expanded over such a non-conformally flat $\mathcal{I}$, like in Eq. (\ref{metforscri}) (or Eq. (\ref{scriaxi}) with axisymmetry) below. Such a linearised theory would then provide a more concrete link between Saw's proposal for the mass/mass-loss formula to those when the gravitational waves are weak.

A paper by Date and Hoque also based their study on the linearised theory \cite{gracos2}, working with two different coordinate systems which are useful and natural according to different viewpoints: 1) Using ``Fermi normal coordinates'' they found corrections to the energy carried by gravitational waves due to the cosmological constant to be in powers of $\Lambda$, just like Saw and Bishop. 2) With ``conformal coordinates'' however, the corrections are in powers of $\sqrt{\Lambda}$, which is a feature of the result reported by Ashtekar et al. \cite{ash3}. This explicitly indicates how different types of coordinate systems would lead to apparently different kinds of corrections. They also defined a gauge-invariant quantity which carries information relating to the polarisation modes of the waves. In relation to this, Bishop similarly argued for the physical relevance of $\Psi_4$ being gauge-invariant and discussed its relationship to the deviation of nearby geodesics using the Riemann tensor.

Hence, a lesson from linearised theory is that whilst they provide useful approximations to the full theory as well as a means for gleaning insightful intuition, care must be taken in making comparisons since different choices of coordinates, tetrads, gauges would drastically lead to seemingly disparate expressions, as emphasised by Saw \cite{Vee2017c}.

\subsection{Asymptotically de Sitter spacetimes that are not consequences of smooth conformal compactification}

Intriguingly, a preprint (at the time of writing this review article) recently appeared \cite{Zhang}, where Xie and Zhang formulated a different boundary condition at infinity for the physical metric. They found that this allows for a different kind of peeling of the Weyl spinor, though the usual peeling can be recovered by introducing certain conditions. A curious result is that for $\Lambda\neq0$, in spite of the absence of the usual Bondi news, there is still a non-zero $O(r^{-1})$ term for $\Psi_4$ due to the cosmological constant. This is an illustration of how the cosmological constant leads to bizarrely new physics of gravitational radiation, which is not found for asymptotically flat spacetimes.

Their metric with such a boundary condition at infinity however, cannot be obtained from the conformal approach. In other words, it would not admit a smooth conformal compactification (as pointed out by Saw \cite{Vee2017c}) and is therefore not equivalent to any of the other work in this area, thus far. This brings up an interesting view and prospect, that perhaps one should consider other kinds of compactification of our physical spacetime, for instance \emph{projective compactification} \cite{Rod} that would lead to physical spacetimes with new properties.

\subsection{The Newman-Unti approach}\label{Section1E}

The rest of this review article is devoted to providing the key results on the asymptotic behaviour of de Sitter-like spacetimes, primarily based on the Newman-Penrose formulation by Saw \cite{Vee2016,Vee2017,Vee2017b,Vee2017c}. The reasons are as follows:
\begin{enumerate}
\item This is an exact result within the \emph{full non-linear} Einstein's theory of general relativity. Many features of asymptotically de Sitter spacetimes as reported by other work/approaches, may be obtained directly using the NP formalism. Such connections will be highlighted and discussed throughout this paper. (The overview given in the previous sub-sections provides a bird's-eye view for the web of relations amongst the raft of papers that have appeared \cite{Vee2016,Vee2017,Vee2017b,Vee2017c,Szabados,Chrusciel,chi1,chi2,ash1,ash2,ash3,ash4,gracos1,gracos2,Zhang}.)
\item This approach has successfully produced the general asymptotic solutions with $\Lambda\in\R$, including the presence of Maxwell fields. Whilst the original aim was to deal with $\Lambda>0$, the results apply even to the case where $\Lambda<0$. This has the implication that one cannot impose conformal flatness on the timelike $\mathcal{I}$ for a universe with $\Lambda<0$, if gravitational waves carry energy away from the source \footnote{This is in sharp contrast to the assumption that conformal flatness of $\mathcal{I}$ in the $\Lambda<0$ case is ``natural'' and serves as a so-called ``reflective boundary condition that can be inserted by hand'' \cite{ash1,ashmag,haw83,reflect1,reflect2}. The ``no Bondi news'' result reported by Ashtekar and Magnon \cite{ashmag} is a repercussion of the stipulation of the timelike $\mathcal{I}$ being conformally flat (albeit with some motivation, e.g. the symmetries of the anti-de Sitter background), just like the $\Lambda>0$ situation \cite{ash1}. See also the recent work by Friedrich that examined the boundary conditions on the timelike $\mathcal{I}$ in studying stability issues with $\Lambda<0$ \cite{Friedrich2014}, and the references found there. Incidentally, Szabados and Tod noted some errors in the analysis carried out by Ref. \cite{ashmag}.}. Furthermore, it \emph{reduces to the known general asymptotic solutions for asymptotically flat spacetimes} (as found in Ref. \cite{Pen88}). This is crucial, because we have so far obtained great agreement between experimental observations and the theory that assumes $\Lambda=0$. Ergo, a generalising theory that includes $\Lambda$ \emph{must be able to reproduce the $\Lambda=0$ results}.
\item The exact expression for outgoing gravitational radiation $\Psi_4$ is obtained, containing the usual shear term $\sigma^o$ as well as new terms due to a non-zero $\Lambda$. Corrections due to a tiny $\Lambda$ may be expanded as a series in powers of $\Lambda$.
\item The mass-loss formula is obtained from one of the Bianchi identities, in a manner analogous to how Newman and Unti got it \cite{newunti62}. A new $\Lambda^2$ term appears with $\Psi_0$, indicating the presence of incoming radiation beyond the isolated system's cosmological horizon that gets picked up by the mass-loss formula, since $\mathcal{I}$ is a spacelike hypersurface when $\Lambda>0$.
\item The structure of $\mathcal{I}$ being non-conformally flat when an isolated gravitating system radiates gravitational waves in a universe with a non-zero $\Lambda$ is manifested. In fact, we have the general expression for the Gauss curvature of these topological 2-spheres of constant $u$ (see Eq. (\ref{Gauss}) in Section \ref{Section2C}), since it arises from one of the NP equations involving the derivatives of $\alpha$ and $\alpha'$ (these are 2 of the 12 complex spin coefficients in the NP formalism, which are essentially the connection coefficients) intrinsic to these 2-surfaces \cite{Pen87}. The general metric for $\mathcal{I}$ can also be obtained (at least, this is analytically carried out explicitly for the axisymmetric case). With this, one can further study the asymptotic symmetries or try solving the Killing equations for $\mathcal{I}$, as discussed below in Section \ref{Section4B}.
\item The peeling properties of the Weyl and Maxwell spinors for asymptotically simple spacetimes are derived in a straightforward way, just like the case where $\Lambda=0$ \cite{newpen62}. Although there are ambiguities in defining the radiation field when $\Lambda\neq0$ \cite{Pen88} (in particular, there is no well-defined timelike infinity $i^+$), the setup here for an isolated system in an electro-$\Lambda$ spacetime naturally places it within a neighbourhood of $r=0$. (This may be seen in the Schwarzschild-de Sitter spacetime, when expressed using spherical coordinates \cite{Vee2016,GP}.) Timelike infinity $i^+$ for this isolated system is $r=0$ when time goes to infinity, and the asymptotic expansion is carried out in powers of $r^{-1}$, where $r\rightarrow\infty$ defines $\mathcal{I}$.
\item As discussed by Penrose \cite{Pen2011}, the null hypersurface $\mathcal{I}$ for $\Lambda=0$ is naturally ruled by a set of null generators that serves as the progression of time. He elaborated that for $\Lambda<0$, the timelike $\mathcal{I}$ similarly allows for a passage of time as discussed by Ashtekar and Magnon \cite{ashmag}. The hypersurface $\mathcal{I}$ being spacelike for $\Lambda>0$ however, implies that any time evolution would bring one away from $\mathcal{I}$, so this somehow makes this case seemingly disparate from the other two.

Nevertheless with an isolated system (which is the physical situation of interest here), we can associate the retarded time coordinate $u$ of the isolated system to define outgoing null cones. For some region sufficiently far away from the isolated system, such outgoing null cones of constant $u$ defined by the isolated system would uniquely construct a foliation of $\mathcal{I}$. One would therefore have a desired description of a ``passage of time on $\mathcal{I}$'' --- as defined by the isolated system.


Furthermore, this then allows for the coordinate $r$ to be taken as an affine parameter of the outgoing null geodesics, and recognise the peeling property of the Weyl (and Maxwell) spinor (as mentioned in the previous point).

One may take such a perspective even in the asymptotically flat case. Then we may think of the $u$ coordinate of the isolated system as defining outgoing null hypersurfaces which provide a foliation of the null $\mathcal{I}$.
\item Incidentally, one may try to define conserved quantities from the relations arising from the Bianchi identities, as was carried out in the asymptotically flat case \cite{newpen65a,newpen68,exton} (although, this may require some kind of generalisation of the usual spherical harmonics on a round 2-sphere to some quantities defined on a \emph{topological 2-sphere}).
\end{enumerate}

This Newman-Unti approach essentially contains most of the results obtained by the conformal studies \cite{Szabados,ash1}, the insights from the linearised gravitational theory and Maxwell theory \cite{ash2,ash3,gracos1,gracos2}, as well as the metric for the non-conformally flat $\mathcal{I}$ for an axisymmetric gravitationally radiating system reported in Ref. \cite{chi1}. On top of that, it also appears to be the only work so far that explicitly yields a ``mass-loss''-type relationship for the \emph{full non-linear} gravitational theory --- arising from one of the Bianchi identities, in the same way the Bondi mass and mass-loss formula were derived for asymptotically flat spacetimes.

In the next section, the preliminaries necessary to express the results and mass-loss formula are given. This begins with an illustration of the spherical coordinates and Newman-Unti null tetrad employed there, by looking at purely vacuum spacetimes with $\Lambda$. With this, the generalisation of the null tetrad to describe asymptotically simple spacetimes is given, in terms of an asymptotic expansion near $\mathcal{I}$. This also requires a stipulation of the fall-offs for the spin coefficients, and it turns out that the peeling property would then follow (without explicitly assuming that $\Psi_0=O(r^{-5})$ when $\Lambda\neq0$) \cite{Vee2017c}. Along the way, we also describe the foliation of $\mathcal{I}$ by topological 2-spheres \cite{Szabados} --- in particular, how it arises that one can only specify \emph{one} such compact 2-surface of constant $u$ to be a round 2-sphere. The explicit metric for $\mathcal{I}$ in the axisymmetric case is given. Perhaps intriguingly, this metric for $\mathcal{I}$ shows how one can obtain Penrose's conclusions that $\mathcal{I}$ is a spacelike, null, or timelike hypersurface, depending on whether $\Lambda>0$, $\Lambda=0$, or $\Lambda<0$ \cite{Pen65}.

Section \ref{Section3} then presents the mass-loss formula for an isolated electro-gravitating system, with the proposed generalisation of the Bondi mass with $\Lambda$ that ensures a positive-definite mass-loss formula. We see how the leading order term of $\Psi_0$ appears explicitly in the mass-loss formula, indicating the presence of incoming radiation from elsewhere that gets picked up since those beyond the cosmological horizon of the isolated system would reach the spacelike $\mathcal{I}$. Further discussions are found in Section \ref{Section4}, where we discuss the Szabados-Tod null tetrad \cite{Vee2017,Szabados} (related to the Newman-Unti null tetrad by a boost transformation), the lack of asymptotic symmetries, as well as conserved quantities, before concluding this review article.

\section{Preliminaries}\label{Section2}

\subsection{Purely vacuum spacetimes}\label{Section2A}

In order to obtain a setup with a cosmological constant $\Lambda$ that reduces to the case where $\Lambda=0$, Saw \cite{Vee2016,Vee2017} employed \emph{spherical coordinates} --- or at least, what would indeed be spherical coordinates for \emph{purely vacuum spacetimes} \cite{GP}. As an illustration, the metrics for purely vacuum spacetimes with $\Lambda\in\R$ are expressed in these coordinates as
\begin{eqnarray}
g=\left(-\frac{\Lambda}{3}r^2+1\right)dt^2-\frac{1}{\left(-\frac{\Lambda}{3}r^2+1\right)}dr^2-r^2d\Omega^2,
\end{eqnarray}
where
\begin{eqnarray*}
\Lambda&=&0\textrm{ is Minkowski spacetime}\\
\Lambda&>&0\textrm{ is de Sitter spacetime}\\
\Lambda&<&0\textrm{ is anti-de Sitter spacetime.}
\end{eqnarray*}
The round unit sphere may be expressed by the usual $\theta$ and $\phi$ spherical angular coordinates, so $d\Omega^2=d\theta^2+\sin^2{\theta}d\phi^2$. These coordinates take values $t\in\R$, $r\in[0,\infty)$, $\theta\in[0,\pi]$, $\phi\in[0,2\pi)$.

Under some coordinate transformation \cite{Vee2016}, this metric can be expressed in terms of a ``retarded null coordinate'' $u$
\begin{eqnarray}\label{puremet}
g=\left(-\frac{\Lambda}{3}r^2+1\right)du^2+2dudr-r^2d\Omega^2,
\end{eqnarray}
so the \emph{outgoing null hypersurfaces} are those where $u$ is constant. The components of the inverse metric are
\begin{eqnarray}\label{puremetinv}
g^{ab}&=&
\begin{pmatrix}
  0 & 1 & 0 & 0 \\
  1 & \frac{\Lambda}{3}r^2-1 & 0 & 0 \\
  0 & 0 & -\frac{1}{r^2} & 0 \\
	0 & 0 & 0 & -\frac{1}{r^2}\csc^2{\theta}
\end{pmatrix}.
\end{eqnarray}
Note that for de Sitter spacetime $\Lambda>0$, there is a cosmological horizon at $r=r_H=\sqrt{3/\Lambda}$ where $r$ being spacelike for $r<r_H$ would become timelike when $r>r_H$. This is expected, since these coordinates cover the entire de Sitter spacetime with $r$ being the usual spherical coordinate radius, as well as $r\rightarrow\infty$ defining null infinity $\mathcal{I}$ --- which is a spacelike hypersurface having a \emph{timelike} vector normal to it (see also the description in section 4.3 of Ref. \cite{GP}). In other words, the hypersurfaces defined by $r=$ constant are timelike when $r<r_H$, null (this is the cosmological horizon) when $r=r_H$, and spacelike when $r>r_H$ (where in particular, $r\rightarrow\infty$ gives the spacelike hypersurface $\mathcal{I}$).

Now, a Newman-Unti null tetrad for the metric in Eq. (\ref{puremet}) may be defined as
\begin{eqnarray}
\vec{l}&=&\vec{\partial}_r\label{NUpure1}\\
\vec{n}&=&\vec{\partial}_u+\left(\frac{\Lambda}{6}r^2-\frac{1}{2}\right)\vec{\partial}_r\\
\vec{m}&=&\frac{1}{\sqrt{2}r}\vec{\partial}_\theta+\frac{i}{\sqrt{2}r}\csc{\theta}\vec{\partial}_\phi\\
\vec{\bar{m}}&=&\frac{1}{\sqrt{2}r}\vec{\partial}_\theta-\frac{i}{\sqrt{2}r}\csc{\theta}\vec{\partial}_\phi,\label{NUpure2}
\end{eqnarray}
where $g^{ab}=l^an^b+n^al^b-m^a\bar{m}^b-\bar{m}^am^b$ relates to the inverse metric components in Eq. (\ref{puremetinv}). Well, $\vec{l}$ points along the outgoing null direction whereas $\vec{n}$ points along the incoming null direction. The former induces a congruence of null geodesics that generates those null hypersurfaces $u=$ constant (the covariant version incidentally, satisfies $\tilde{l}=\tilde{d}u$), and the coordinate $r$ is in this case an affine parameter for those null geodesics. The complex vectors $\vec{m}$ and $\vec{\bar{m}}$ would span compact 2-surfaces containing all the null generators. These null tetrad vectors satisfy the usual orthonormalisation conditions \cite{newpen62}. (Have a look at Fig. \ref{fig1}, though that picture is for general asymptotically de Sitter spacetimes instead of just purely vacuum de Sitter spacetime \footnote{Having such a family of null hypersurfaces is always possible in a normal hyperbolic Riemannian manifold, which was applied to asymptotically flat spacetimes by Newman and Unti \cite{newunti62}. Sachs did something similar and obtained a general form of the metric for asymptotically flat spacetimes \cite{Sachs62}.}.)

\subsection{Asymptotically simple spacetimes}\label{Section2B}

Based on the Newman-Unti null tetrad for purely vacuum spacetimes in Eqs. (\ref{NUpure1})-(\ref{NUpure2}), a generalisation to \emph{asymptotically simple spacetimes} is \cite{Vee2016}
\begin{eqnarray}
\vec{l}&=&\vec{\partial}_r\label{NU1}\\
\vec{n}&=&\vec{\partial}_u+U\vec{\partial}_r+X^\mu\vec{\partial}_\mu\\
\vec{m}&=&\omega\vec{\partial}_r+\xi^\mu\vec{\partial}_\mu\\
\vec{\bar{m}}&=&\bar{\omega}\vec{\partial}_r+\overline{\xi^\mu}\vec{\partial}_\mu,\label{NU2}
\end{eqnarray}
where $\mu$ denotes the two coordinates $\theta$ and $\phi$ which are labels for the null generators of those outgoing null hypersurfaces $u=$ constant, and in general do not correspond to the usual two spherical angular coordinates. The functions $U(u,r,\theta,\phi)$, $X^\mu(u,r,\theta,\phi)$, $\omega(u,r,\theta,\phi)$, $\xi^\mu(u,r,\theta,\phi)$ are prescribed to have the following expansions in inverse powers of $r$ with sufficiently many orders \footnote{For detailed elaborations on how these fall-offs as well as those for the spin coefficients in Eqs. (\ref{SC01})-(\ref{SC02}) were stipulated and subsequently found to be consistent with those obtained by studying the conformal structure from Ref. \cite{Szabados}, please have a look at Ref. \cite{Vee2016}.}
\begin{eqnarray}
U&=&\frac{\Lambda}{6}r^2+O(1)\label{U}\\
X^\mu&=&O(r^{-1})\\
\omega&=&O(r^{-1})\\
\xi^\mu&=&(\xi^\mu)^or^{-1}+O(r^{-2}).\label{defxi}
\end{eqnarray}
The coefficients of each term in the expansions are unknown functions of $u,\theta,\phi$, which are to be determined by solving the 38 vacuum NP equations \cite{Vee2016} (or 38 NP equations with Maxwell fields, plus the 4 Maxwell equations \cite{Vee2017}).

Before solving the NP equations, the fall-offs for the 12 complex spin coefficients are stipulated. Some geometrical properties like $\vec{l}$ being tangent to the outgoing null geodesics, the remaining null tetrad vectors being parallel transported along $\vec{l}$, the freedoms involving the origin of the affine parameter $r$ of those outgoing null geodesics, as well as spatial rotations relating to $\vec{m}$ and $\vec{\bar{m}}$ are applied, and the fall-offs are
\begin{eqnarray}\label{SC01}
\kappa&=&0, \gamma'=0, \tau'=0,\\
\kappa'&=&O(1)\\
\sigma&=&O(r^{-1})\\
\sigma'&=&O(1)\\
\tau&=&O(r^{-1})\\
\gamma&=&\gamma^o_{-1}r+O(r^{-2})\\
\rho&=&\rho^o_1r^{-1}+O(r^{-3})\\
\rho'&=&\rho'^o_{-1}r+O(r^{-1})\\
\alpha&=&O(r^{-1})\\
\alpha'&=&O(r^{-1}).\label{SC02}
\end{eqnarray}
These were stated by studying the Schwarzschild-de Sitter spacetime (which implies that $\rho'^o_{-1}$ and $\gamma^o_{-1}$ are non-zero due to a cosmological constant $\Lambda$), as well as to ensure that $\sigma^o$ (which is the $r^{-2}$ term for $\sigma$) is not forced to vanish (and this gives rise to $O(1)$ terms for $\sigma'$ and $\kappa'$) \footnote{Incidentally, the integration function $\sigma^o(u,\theta,\phi)$ is freely-specifiable even with a cosmological constant present. Since this is the \emph{asymptotic shear} \cite{ANK}, all asymptotically simple spacetimes allow for the existence of \emph{asymptotically shear-free null congruences}.}. Note that such a stipulation of the fall-offs in Eqs. (\ref{U})-(\ref{SC02}) corresponds to fixing the various gauge freedoms, and the asymptotic solutions obtained in Refs. \cite{Vee2016,Vee2017} are \emph{general}.

The final piece of required information is to have the fall-off for the dyad component of the Weyl spinor $\Psi_0$ being $O(r^{-5})$ \footnote{It was lately found that the additional requirement ``$\Psi_0=O(r^{-5})$'' is actually not necessary when there is a \emph{non-zero} $\Lambda$. The stipulation of the fall-offs for the spin coefficients and the null tetrad together with a non-zero $\Lambda$ \emph{will imply} that $\Psi_0=O(r^{-5})$ \cite{Vee2017c}.}. The asymptotic solutions are then found \cite{Vee2016}, and extended to include the Maxwell fields \cite{Vee2017} \footnote{Similarly, the dyad component of the Maxwell spinor $\phi_0$ is prescribed to have a fall-off of $O(r^{-3})$ in Ref. \cite{Vee2017}, but it was later realised that this is not necessary if $\rho=-r^{-1}+O(r^{-3})$.}. Note that setting $\Lambda=0$ reduces these results for asymptotically simple spacetimes to the well-known asymptotic solutions for asymptotically flat empty spacetimes in Ref. \cite{newunti62} (or Ref. \cite{Pen88} including Maxwell fields).

\subsection{Foliation of $\mathcal{I}$ by topological 2-surfaces}\label{Section2C}

For asymptotically flat spacetimes, one can choose the leading order terms $(\xi^{\mu})^o$ in Eqs. (\ref{defxi}) such that they correspond to a round unit sphere \cite{newunti62}. With a cosmological constant however, this cannot be assumed in general \cite{Szabados}. If such a choice is insisted, that would lead to a conformally flat $\mathcal{I}$ with the ramification that gravitational waves do not carry energy away from the isolated source \cite{ash1}. In the NP formalism, this is reflected by the pair of metric equations involving $D'\xi^\mu$, which yield \cite{Vee2016}
\begin{eqnarray}\label{xi}
(\dot{\xi}^\mu)^o=-\frac{\Lambda}{3}\sigma^o(\overline{\xi^\mu})^o,
\end{eqnarray}
where the dot is derivative with respect to $u$.

\begin{figure}
\centering
\includegraphics[width=12.5cm]{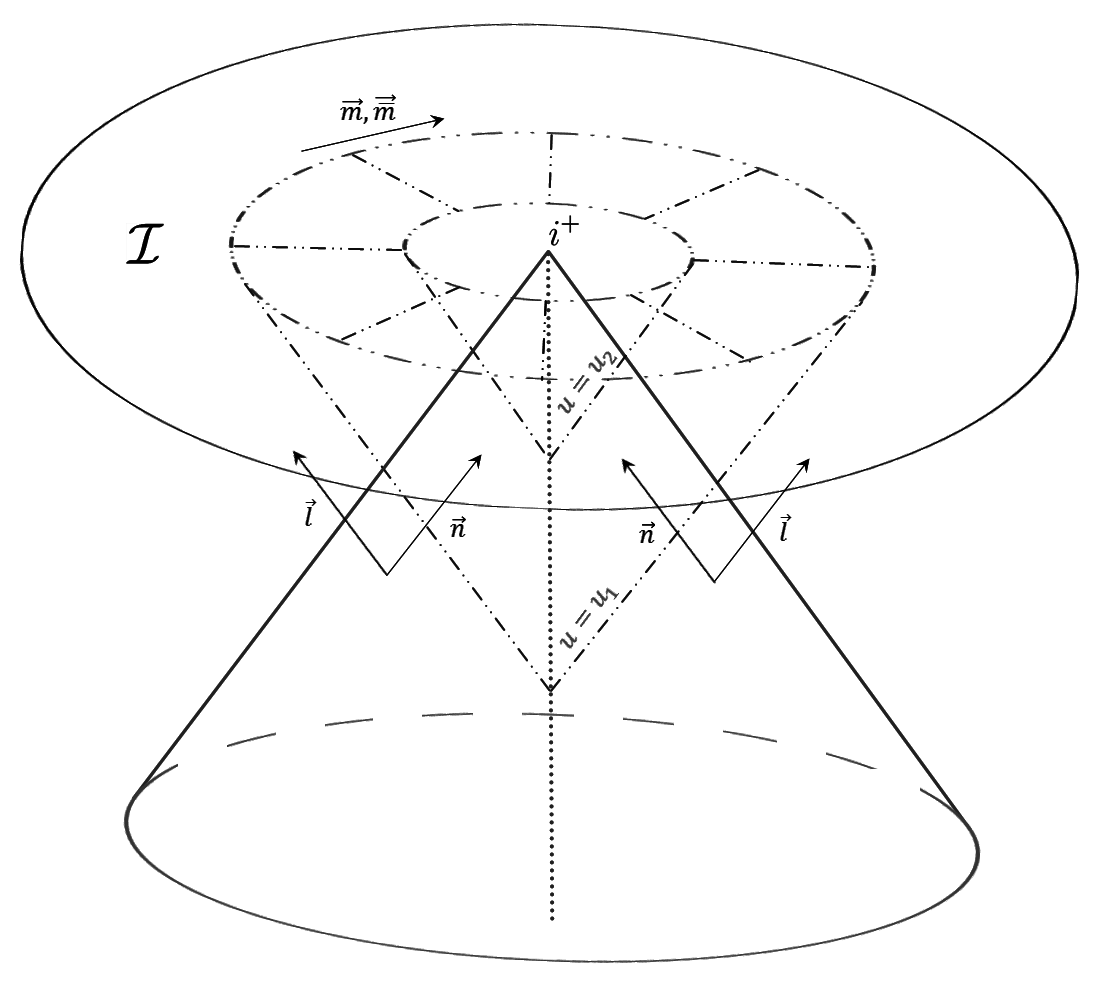}
\caption{Penrose diagram for an isolated system with $\Lambda>0$, near future null infinity $\mathcal{I}$. The vertical dotted line is $r=0$, representing the timelike worldline of the \emph{compact isolated system}. The isolated system is situated at (or in a neighbourhood of) $r=0$, with timelike infinity $i^+$ and a cosmological horizon\footnote{For the Schwarzschild-de Sitter spacetime in these spherical coordinates \cite{Vee2016,GP}, it is easy to see that the black hole is located at the origin where $r=0$. Whilst timelike infinity $i^+$ is well-defined for $\Lambda=0$, the presence of $\Lambda$ makes this origin-dependent. This origin-dependence also results in the concept of the ``radiation field'' being not very well-defined \cite{Pen88}. Nevertheless, within the context of studying an isolated system, we \emph{do} have a natural choice for the origin $r=0$.}. Each value of $u=$ constant defines an outgoing null hypersurface whose apex lies on the vertical dotted line. Their intersections with the spacelike hypersurface null infinity $\mathcal{I}$ are compact 2-surfaces, thereby providing a foliation of $\mathcal{I}$ by these compact 2-surfaces of constant $u$. When $\sigma^o\neq0$, these compact 2-surfaces of constant $u$ on $\mathcal{I}$ are necessarily \emph{topological 2-spheres}. The Newman-Unti null tetrad vectors $\vec{l}$ and $\vec{n}$ are outgoing and incoming null vectors, respectively. The remaining $\vec{m}$ and $\vec{\bar{m}}$ span those compact 2-surfaces. The coordinate $r$ is also an affine parameter for the null geodesics which generate those null hypersurfaces of constant $u$.}
\label{fig1}
\end{figure}

Well, $\mathcal{I}$ can be thought of as being foliated by these 2-surfaces of constant $u$ (see the setup of $\mathcal{I}$ with $\Lambda>0$ in section 4.2 of Ref. \cite{Szabados} that describes this). The above equations mean that whilst one may pick one such 2-surface on $\mathcal{I}$ to be a round sphere for some $u=u_0$, a different $u$ would generally have a \emph{topological 2-sphere} when $\Lambda\neq0$ \emph{and} $\sigma^o\neq0$. Furthermore, the Gauss curvature for this topological 2-sphere is
\begin{eqnarray}\label{Gauss}
K=1+\frac{2\Lambda}{3}\int{\textrm{Re}(\eth^2\bar{\sigma}^o)du},
\end{eqnarray}
obtained from the NP equation involving the derivatives of $\alpha$ and $\alpha'$ intrinsic to those 2-surfaces \cite{Vee2016,Pen87}. Here, $\eth$ (as well as $\eth'$) is a derivative operator intrinsic to these 2-surfaces, defined in an expected manner and acts on spin-weighted quantities (see the summary of the asymptotic solutions with $\Lambda$ in Ref. \cite{Vee2016} and/or Ref. \cite{Vee2017}). If we have $\sigma^o=0$, then $\mathcal{I}$ is conformally flat --- its Cotton-York tensor is zero. With the right-hand side of Eqs. (\ref{xi}) being 0 (due to $\sigma^o=0$), the choice of $(\xi^o)^\mu$ being a round sphere at some $u=u_0$ would remain even as $u$ varies. Also, we see that the Gauss curvature is just 1. The Penrose diagram in Fig. \ref{fig1} illustrates the geometrical setup for asymptotically de Sitter spacetimes.

\subsubsection{Axisymmetry}\label{Section2C1}

We can glean some insights explicitly by considering an axisymmetric system (i.e. no $\phi$ dependence), since Eqs. (\ref{xi}) may then be solved analytically \footnote{As noted by Sachs \cite{Sachs62}, the full treatment for a general system does not offer new physics but only makes the technicalities so much more complicated: ``\emph{... in most arguments, we can confine our attention to the axially symmetric case without any essential loss of generality.}''}. With $\displaystyle3f(u,\theta):=\int{\sigma^o(u,\theta)du}$ where $\sigma^o(u,\theta)$ is real \footnote{Well, the two polarisation modes of gravitational waves are encoded into the real and imaginary parts of $\sigma^o$. With axisymmetry, there is only one polarisation mode \cite{ash4} and we can thus specify $\sigma^o(u,\theta)$ to be a real function.}, we find that
\begin{eqnarray}
(\xi^\theta)^o&=&\frac{1}{\sqrt{2}}e^{-\Lambda f(u,\theta)}\\
(\xi^\phi)^o&=&\frac{i}{\sqrt{2}}e^{\Lambda f(u,\theta)}\csc{\theta}
\end{eqnarray}
satisfy that pair of equations, and therefore the metric for the 2-surface of constant $u$ on $\mathcal{I}$ is
\begin{eqnarray}\label{2axi}
g_{2,axi}=e^{2\Lambda f(u,\theta)}d\theta^2+e^{-2\Lambda f(u,\theta)}\sin^2{\theta}d\phi^2.
\end{eqnarray}
Notice that this is not a round unit sphere, unless $\Lambda=0$ (i.e. spacetime is asymptotically flat), or $f=0$ (so $\sigma^o=0$, such that outgoing gravitational waves carry no energy away from the system).

The Gauss curvature for this axisymmetric topological 2-sphere and the action of the $\eth'$ operator on $\sigma^o$ have been worked out explicitly in section 6 of Ref. \cite{Vee2016}. They both have overall factors involving $e^{-\Lambda f}$. As these appear in the generalised mass in Eq. (\ref{topsph}) and the generalised mass-loss formula in Eq. (\ref{0Bondimasslosspsi0ST}) to include $\Lambda>0$ (in Section \ref{Section3}), one sees that a Taylor expansion of the exponential factor gives \emph{terms made up of all positive integer powers of $\Lambda$ as corrections due to the cosmological constant}.

\subsection{Character of null infinity $\mathcal{I}$ due to $\Lambda$}\label{Section2D}

The full general (without axisymmetry) $(3+1)$-$d$ spacetime would have the metric
\begin{eqnarray}
g=-\frac{\Lambda}{3}r^2du^2-r^2g_2+O(r),
\end{eqnarray}
where $g_2$ is the metric for the topological 2-sphere constructed from $(\xi^\mu)^o$, and $O(r)$ represents the collection of terms of higher order than $r^2$, i.e. $r^{-2}$ times those terms represented by $O(r)$ would go to zero as $r\rightarrow\infty$. Now, we may define a conformally rescaled metric $\tilde{g}:=C^2g$, where $C=r^{-1}$ is the conformal factor. Therefore,
\begin{eqnarray}
\tilde{g}=-\frac{\Lambda}{3}du^2-g_2+O(r^{-1}),
\end{eqnarray}
and in the limit where $r\rightarrow\infty$, this gives the metric for null infinity $\mathcal{I}$ \footnote{Recall that the hypersurfaces $u=$ constant are outgoing null cones, with $r$ being an affine parameter of the null geodesics generating these null cones. So $r\rightarrow\infty$ would reach infinity along a null light ray, which is null infinity --- since by definition, null infinity is the hypersurface where light rays would eventually end up after infinite physical time.}
\begin{eqnarray}\label{metforscri}
\tilde{g}=-\frac{\Lambda}{3}du^2-g_2.
\end{eqnarray}
With axisymmetry, we have $g_2=g_{2,axi}$ in Eq. (\ref{2axi}) so
\begin{eqnarray}\label{scriaxi}
\tilde{g}=-\frac{\Lambda}{3}du^2-e^{2\Lambda f(u,\theta)}d\theta^2-e^{-2\Lambda f(u,\theta)}\sin^2{\theta}d\phi^2.
\end{eqnarray}

We see that the nature of the hypersurface $\mathcal{I}$ depends on the cosmological constant: It is a \emph{spacelike hypersurface} for $\Lambda>0$, a \emph{timelike hypersurface} for $\Lambda<0$, and \emph{degenerates into a null hypersurface} for $\Lambda=0$. This is an equivalent way of stating Penrose's conclusions of the character of $\mathcal{I}$, where he obtained an expression for the squared length of the vector normal to $\mathcal{I}$ and found it to be proportional to $\Lambda$ \cite{Pen65}. Here, we have a complementary point of view from the metric for $\mathcal{I}$ itself (instead of the vector normal to $\mathcal{I}$) involving the cosmological constant which directly determines its character.

Incidentally, the axisymmetric metric for $\mathcal{I}$ given by Eq. (\ref{scriaxi}) is also found by He and Cao \cite{chi1} who studied this problem of gravitational radiation with a cosmological constant by employing Bondi et al.'s original approach \cite{Bondi62} in enunciating an ansatz for the axisymmetric metric. To include the presence of $\Lambda$, they postulated a new leading term in the function $\gamma$ (which when prescribed for some $u$, allows for other unknowns to be solved systematically via the Einstein field equations). This new leading term in their ansatz would also lead to $\mathcal{I}$ with a non-vanishing Cotton-York tensor, i.e. Eq. (\ref{scriaxi}). Note that whilst outgoing gravitational radiation would alter the structure of $\mathcal{I}$ (as manifested by a non-zero $\sigma^o$), Maxwell/electromagnetic radiation on the other hand has no effect on $\mathcal{I}$ \cite{Vee2017,chi2}.

\section{Mass-loss formula with $\Lambda>0$}\label{Section3}

For asymptotically flat spacetimes, the mass-loss formula arises from the Bianchi identity involving $D'\Psi_2$. This gives a $u$-derivative of a quantity related to the mass aspect, viz. $-(\Psi^o_2+\sigma^o\dot{\bar{\sigma}}^o)$, where $\Psi^o_2$ is the leading order term of the dyad component of the Weyl spinor $\Psi_2$ when expanded as inverse powers of $r$. The superscript $^o$ denotes that this is a function of $u,\theta,\phi$, i.e. it is independent of $r$. This relationship has no $r$ dependence, as the $r^{-2}$ factors all cancel out. With the presence of $\Lambda$, we find that the corresponding relationship is \cite{Vee2017}
\begin{eqnarray}\label{masslosspreintegration}
-\frac{\partial}{\partial u}(\Psi^o_2+\sigma^o\dot{\bar{\sigma}}^o)&=&-|\dot{\sigma}^o|^2-\eth\Psi^o_3-k|\phi^o_2|^2-\frac{\Lambda}{3}K|\sigma^o|^2+\frac{\Lambda}{3}\sigma^o\eth'\eth\bar{\sigma}^o+\frac{\Lambda}{6}\eth'\Psi^o_1-\frac{2\Lambda^2}{9}|\sigma^o|^4\nonumber\\&&-\frac{\Lambda^2}{18}\textrm{Re}(\bar{\sigma}^o\Psi^o_0)+\frac{k\Lambda^2}{36}|\phi^o_0|^2.
\end{eqnarray}
Just like $\Psi^o_2$, the quantities $\Psi^o_0$, $\Psi^o_1$, $\Psi^o_3$, $\Psi^o_4$ are leading order terms of $\Psi_0$, $\Psi_1$, $\Psi_3$, $\Psi_4$ when expanded as inverse powers of $r$, respectively. Similarly, $\phi^o_0$, $\phi^o_1$, $\phi^o_2$ are leading order terms of the dyad components of the Maxwell spinor $\phi_0$, $\phi_1$, $\phi_2$, respectively, and $k=2G/c^4$ is the coupling constant of the stress-energy tensor to the Maxwell fields.

For asymptotically flat spacetimes, the Bondi mass is $\displaystyle M_B=-\frac{1}{A}\oint{(\Psi^o_2+\sigma^o\dot{\bar{\sigma}}^o)}d^2S$, where integration is over a compact 2-surface of constant $u$ on $\mathcal{I}$ and $A$ is the area \cite{newunti62}. A proposed generalisation by Saw in Ref. \cite{Vee2016} that includes $\Lambda>0$ is
\begin{eqnarray}\label{topsph}
M_{\Lambda}:=M_B+\frac{1}{A}\int{\left(\oint{\left(\Psi^o_2+\sigma^o\dot{\bar{\sigma}}^o\right)\frac{\partial}{\partial u}(d^2S)}\right)du}+\frac{\Lambda}{3A}\int{\left(\oint{K|\sigma^o|^2d^2S}\right)du},
\end{eqnarray}
which ensures that this mass $M_\Lambda$ strictly decreases whenever $\sigma^o$ is non-zero in the absence of incoming gravitational (and electromagnetic) radiation, i.e. the mass of an isolated (electro-)gravitating system strictly decreases due to energy carried away by gravitational (and electromagnetic) waves \cite{Vee2016,Vee2017}. Note that the second term on the right-hand side of Eq. (\ref{topsph}) arises due to the topological 2-sphere's surface element $d^2S$ having a $u$-dependence when $\Lambda\neq0$, together with a process of interchanging the order of taking a $u$-derivative and integrating over the topological 2-sphere. Now, integration of Eq. (\ref{masslosspreintegration}) over a compact 2-surface of constant $u$ on $\mathcal{I}$ gives the sought after \emph{generalisation of the mass-loss formula with $\Lambda>0$} \footnote{If $\Lambda<0$, then this form of the mass-loss formula does not ensure that the mass $M_\Lambda$ strictly decreases due to energy carried away by outgoing gravitational radiation. This is due to the term involving the factor $\Lambda$ now having an opposite sign. In that case, one may propose that the definition of the mass in Eq. (\ref{topsph}) absorbs this term as well.}
\begin{eqnarray}
\frac{dM_{\Lambda}}{du}
&=&-\frac{1}{A}\oint{\left(|\dot{\sigma}^o|^2+k|\phi^o_2|^2+\frac{\Lambda}{3}|\eth'\sigma^o|^2+\frac{2\Lambda^2}{9}|\sigma^o|^4+\frac{\Lambda^2}{18}\textrm{Re}(\bar{\sigma}^o\Psi^o_0)-\frac{k\Lambda^2}{36}|\phi^o_0|^2\right)d^2S},\nonumber\\\label{0Bondimasslosspsi0ST}
\end{eqnarray}
where the ``divergence terms'' involving $\eth'\Psi^o_1$ and $\eth\Psi^o_3$ vanish upon integration over the compact 2-surface which has no boundary. Also, integration by parts have been applied on the term with $\sigma^o\eth'\eth\bar{\sigma}^o$. We may immediately observe that with the presence of $\Lambda$, then $\sigma^o$ itself would contribute to the mass-loss formula. (Recall that for asymptotically flat spacetimes in Eq. (\ref{Bondimasslossflat}), one needs a variation of $\sigma^o$ with $u$ to give rise to a mass loss.)

If there are no Maxwell fields present, the mass-loss formula purely due to gravitation is
\begin{eqnarray}
\frac{dM_{\Lambda}}{du}
&=&-\frac{1}{A}\oint{\left(|\dot{\sigma}^o|^2+\frac{\Lambda}{3}|\eth'\sigma^o|^2+\frac{2\Lambda^2}{9}|\sigma^o|^4+\frac{\Lambda^2}{18}\textrm{Re}(\bar{\sigma}^o\Psi^o_0)\right)d^2S}.\label{0BondimasslossGraonly}
\end{eqnarray}

If there are only Maxwell fields but no gravitational radiation, i.e. $\sigma^o=0$, then
\begin{eqnarray}
\frac{dM_{\Lambda}}{du}
&=&-\frac{k}{A}\oint{\left(|\phi^o_2|^2-\frac{\Lambda^2}{36}|\phi^o_0|^2\right)d^2S},\label{0BondimasslossMaxonly}
\end{eqnarray}
where $M_\Lambda=M_B$ (since the compact 2-surface of constant $u$ on $\mathcal{I}$ is a round sphere when $\sigma^o=0$ --- see Eq. (\ref{Gauss}), hence the surface element $d^2S$ does not have a $u$-dependence). In other words, unlike gravity the Maxwell fields do not alter the structure of $\mathcal{I}$ and $M_\Lambda$ is just the same as $M_B$ for asymptotically flat spacetimes.

Incidentally, the Maxwell equation involving $D'\phi_1$ gives rise to one relation which is independent of $r$ (as these factors of $r^{-2}$ cancel out)
\begin{eqnarray}
\dot{\phi^o_1}=\eth\phi^o_2-\frac{\Lambda}{6}\eth'\phi^o_0.
\end{eqnarray}
Integrating this over a compact 2-surface of constant $u$ implies \emph{total charge conservation}, i.e. the term proportional to $\displaystyle\oint{\phi^o_1d^2S}$ (plus a term analogous to the second term on the right-hand side of Eq. (\ref{topsph})) does not depend on $u$.

\subsection{Incoming radiation}\label{Section3A}

\begin{figure}
\centering
\includegraphics[width=15.85cm]{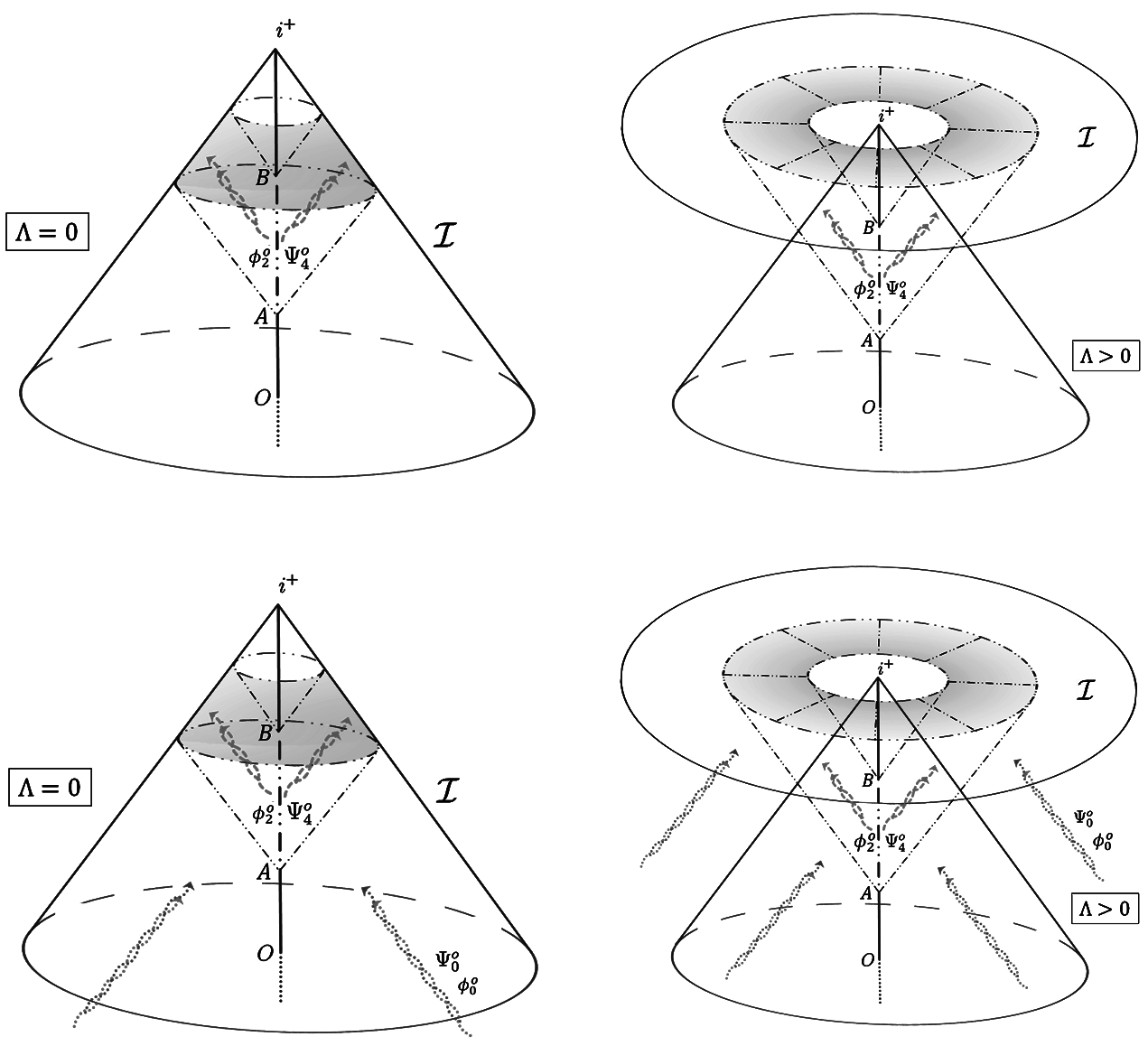}
\caption{Penrose diagrams for an isolated electro-gravitating system emitting gravitational and/or electromagnetic radiation $\Psi^o_4$, $\phi^o_2$ in an asymptotically flat (top left) and asymptotically de Sitter (top right) spacetimes. The isolated system emits gravitational/electromagnetic radiation between $A$ and $B$, and the radiation would eventually hit $\mathcal{I}$ over the shaded region. If incoming radiation $\Psi^o_0$, $\phi^o_0$ from elsewhere is present (bottom figures), then those waves beyond the cosmological horizon of the isolated system for the case with $\Lambda>0$ would also hit the spacelike $\mathcal{I}$ and get picked up by the mass-loss formula in Eq. (\ref{0Bondimasslosspsi0ST}).}
\label{fig2}
\end{figure}

The generalised mass-loss formula with $\Lambda$ in Eq. (\ref{0Bondimasslosspsi0ST}) contains unusual terms having $\Psi^o_0$ and $\phi^o_0$ coupled via a $\Lambda^2$ factor (so the overall sign is the same regardless of the sign of $\Lambda\neq0$). Well, these terms may be interpreted as incoming gravitational radiation and electromagnetic radiation, respectively (these are radiation along the direction of $\vec{n}$), just as $\Psi^o_4(\sigma^o)$ and $\phi^o_0$ are interpreted as outgoing ones (these are radiation along the direction of $\vec{l}$) \footnote{Well, the outgoing gravitational radiation $\Psi^o_4(\sigma^o)$ is a function of $\sigma^o$. In addition, $\Psi^o_4$ itself also contains a term $\Lambda^2\bar{\Psi}^o_0/36$ \cite{Vee2016,Vee2017}.}. The presence of a $\Lambda>0$ (corresponding to describing our universe which expands at an accelerated rate) implies that $\mathcal{I}$ is now spacelike, instead of null when $\Lambda=0$. Hence, if there is incoming radiation from elsewhere, then those beyond the isolated system's cosmological horizon (which cannot influence it) would reach $\mathcal{I}$ and get picked up by the mass-loss formula in Eq. (\ref{0Bondimasslosspsi0ST}) (see Fig. \ref{fig2}). This is perspicuous for the case of electromagnetism in Eq. (\ref{0BondimasslossMaxonly}): $|\phi^o_2|^2$ strictly decreases $M_\Lambda$ whilst $|\phi^o_0|^2$ strictly increases $M_\Lambda$. This is however, rather strange for the gravitational version in Eq. (\ref{0BondimasslossGraonly}). In particular, if the system does not emit gravitational radiation or $\sigma^o=0$, then the incoming gravitational radiation from elsewhere would not register with the mass-loss formula. In fact, only the real part of $\bar{\sigma}^o\Psi^o_0$ is recognised and even so, the sign is not quite definite.

Ref. \cite{ash2} that dealt with the linearised gravitational theory and Maxwell fields with $\Lambda>0$ (then subsequently went on to work out the quadrupole formula in Ref. \cite{ash3}), found that it is possible for the energy of both gravitational and electromagnetic waves emitted by the isolated system to be \emph{arbitrarily negative}. They nevertheless explained that this observation corresponds to the timelike Killing field (within the isolated system's cosmological horizon) becoming spacelike and \emph{past directed} outside the isolated system's cosmological horizon. For an isolated electro-gravitating system without any incoming radiation from elsewhere, these negative contributions are zero so the energy carried away from the system due to the emitted radiation is \emph{necessarily positive-definite}.

The condition for \emph{no incoming radiation} corresponds to setting $\Psi^o_0(u,\theta,\phi)$ ($\phi^o_0(u,\theta,\phi)$ for Maxwell) to zero. Well, one of the Bianchi identities involving $D'\Psi_0$ gives a relationship for $\dot{\Psi^o_0}$ at order $r^{-5}$ (where all factors of $r^{-5}$ cancel out), viz. $\dot{\Psi}^o_0=\eth\Psi^o_1+3\sigma^o\Psi^o_2+\Lambda\Psi^1_0/6$ \cite{Vee2016}. This means that one is only allowed to specify $\Psi^o_0$ on some initial hypersurface $u=u_0$. In the asymptotically flat case where $\Lambda=0$, setting $\Psi^o_0=0$ would hence introduce a constraint between $\eth\Psi^o_1$ and $3\sigma^o\Psi^o_2$. Remarkably, a non-zero $\Lambda$ introduces a new free function, namely $\Psi^1_0(u,\theta,\phi)$ which is the next order term of $\Psi_0$ (when expanded in inverse powers of $r$). The analogous situation holds for $\phi^o_0$ and the Maxwell equation involving $D'\phi_0$ \cite{Vee2017}.

To summarise, we see that ``no incoming radiation'' means that $\Psi_0$ [or $\phi_0$] has to have a weaker fall-off of $O(r^{-6})$ [or $O(r^{-4})$] instead of $O(r^{-5})$ [or $O(r^{-3}$)], such that the radiation is sufficiently negligible beyond the isolated system's cosmological horizon and would not carry information all the way to the spacelike $\mathcal{I}$. Incidentally, the mass-loss formula for an isolated electro-gravitating system without any incoming radiation from elsewhere is
\begin{eqnarray}
\frac{dM_{\Lambda}}{du}
&=&-\frac{1}{A}\oint{\left(|\dot{\sigma}^o|^2+k|\phi^o_2|^2+\frac{\Lambda}{3}|\eth'\sigma^o|^2+\frac{2\Lambda^2}{9}|\sigma^o|^4\right)d^2S},\label{Bondimassloss}
\end{eqnarray}
so the energy carried away by the outgoing radiation is manifestly positive-definite --- which is the basis for the definition of $M_\Lambda$ in Eq. (\ref{topsph}).

\section{Discussion}\label{Section4}

\subsection{The peeling property and the Szabados-Tod null tetrad}\label{Section4A}

The \emph{peeling property} of the Weyl curvature for asymptotically \emph{simple} (i.e. including a cosmological constant) spacetimes has been shown by Roger Penrose \cite{Pen65,Pen88}, involving direct use of spinors and the properties of conformally rescaled spacetimes. This generalised Sachs's original result for asymptotically \emph{flat} spacetimes \cite{Sachs61} \footnote{See also the references listed in Ref. \cite{Sachs61} for some developments by others prior to the work by Sachs.}. Prior to the use of conformal techniques and explicit application of spinors, Newman and Penrose were able to get this result using their NP formalism on the physical spacetime for the case where $\Lambda=0$, by assuming that $\Psi_0=O(r^{-5})$ (as well as some technicalities regarding differentiability conditions) \cite{newpen62}. The Bianchi identities involving $D\Psi_n$ for $n=1,2,3,4$ possess a nice hierarchical structure allowing for a systematic integration procedure to yield $\Psi_n=O(r^{n-5})$ --- which is the peeling property.

With a cosmological constant $\Lambda$, Saw \cite{Vee2016} made the same starting assumptions like $\Psi_0=O(r^{-5})$ and obtained the same result. This works because $\Lambda$ does not affect the Ricci spinor and being a constant, the Ricci scalar (which is proportional to $\Lambda$) is also a constant \cite{Don}. What appear in those Bianchi identities are \emph{derivatives} of the Ricci scalar (as well as the derivatives of the Ricci spinor), and therefore the presence of a cosmological constant has no effect --- thereby we have the peeling property even with $\Lambda\neq0$. Recently, it was noted that with a non-zero $\Lambda$, there is no need to assume the condition $\Psi_0=O(r^{-5})$ as this would follow from the stipulation of the fall-offs for the spin coefficients and the unknown functions in the null tetrad \cite{Vee2017c}. A similar peeling property for the Maxwell fields also holds \cite{Vee2017}.

The results by Saw were presented using the Newman-Unti null tetrad, following their approach for the asymptotically flat case \cite{newunti62}. For asymptotically de Sitter spacetimes, work by Szabados and Tod \cite{Szabados} using the conformally rescaled de Sitter-like spacetimes employed a different $\vec{l}$ and $\vec{n}$ \footnote{They also used different $\vec{m}$ and $\vec{\bar{m}}$ vectors which do not have a $\vec{\partial}_r$ component. On the other hand, those in Refs. \cite{newunti62,Vee2016} are set to be parallel transported along $\vec{l}$.}, corresponding to a symmetric scaling of the conformal factors over the spinor dyads $o^A$ and $\iota^A$. They reasoned that since $\mathcal{I}$ is spacelike, then there is no preferred null direction --- unlike the asymptotically flat case where the null $\mathcal{I}$ specifies a natural null direction. Anyway, the asymptotic solutions to the Einstein-Maxwell-de Sitter spacetimes are presented by Saw \cite{Vee2017} in both the Newman-Unti (NU) and Szabados-Tod (ST) null tetrads, where the latter null tetrad is defined as
\begin{eqnarray}
\vec{l}_{ST}&=&r\vec{l}_{NU}\\
\vec{n}_{ST}&=&\frac{1}{r}\vec{n}_{NU}\\
\vec{m}_{ST}&=&\vec{m}_{NU}\\
\vec{\bar{m}}_{ST}&=&\vec{\bar{m}}_{NU}.
\end{eqnarray}
The subscript $_{NU}$ refers to the Newman-Unti null tetrad defined in Eq. (\ref{NU1})-(\ref{NU2}). These two null tetrads are related by a boost transformation, which affects the fall-offs of various quantities (see Appendix A in Ref. \cite{Vee2017}). In particular, whilst the peeling properties of the Weyl and Maxwell spinors are exhibited in the NU null tetrad, the fall-offs in the ST null tetrad are $\Psi^o_{i}=O(r^{-3})$ and $\phi^o_j=O(r^{-2})$ respectively, where $i=0,1,2,3,4$ and $j=0,1,2$. Note also that the coordinate $r$ in the ST null tetrad is not an affine parameter for the congruence of null geodesics generating the outgoing null hypersurfaces of constant $u$. This is manifested by a non-zero $\gamma'=-1/2$. Furthermore, the spin coefficients $\rho'$ and $\gamma$ in the ST null tetrad do not have leading order terms being $O(r)$ (see Eqs. (\ref{SC01})-(\ref{SC02}) above for the fall-offs using the NU null tetrad) which would appear to ``blow up'' near $\mathcal{I}$ where $r\rightarrow\infty$. Instead, they are of order $O(1)$. The results expressed in both the NU and ST null tetrads (in fact, in any such null tetrads obtained by a boost transformation of NU) would hold for asymptotically flat spacetimes as a special case, when $\Lambda$ is set to zero.

Perhaps most significantly, in spite of the shifting of terms due to different factors of $r$ in the asymptotic expansions near $\mathcal{I}$ using different null tetrads, all the relationships that do not involve $r$ would \emph{remain invariant}. The Bianchi identity in Eq. (\ref{masslosspreintegration}) which gives the mass-loss formula is one such \emph{null-tetrad-independent} relationship. This echoes' Bondi et al.'s \cite{Bondi62} remark that the mass-loss formula is an \emph{exact result} obtained from going towards null infinity $\mathcal{I}$.

\subsection{Asymptotic symmetries}\label{Section4B}

\subsubsection{Asymptotically flat spacetimes}\label{Section4B1}

In the original work by Bondi et al. for asymptotically flat spacetimes, they studied the permissible coordinate transformations that leave their axisymmetric metric ansatz unchanged \cite{Bondi62}. Sachs later treated the general case without axisymmetry \cite{Sachs62}. This set of transformations is commonly referred to as the Bondi-Metzner-Sachs (BMS) group \cite{Pen63}, which is larger than the Poincar\'{e} group and contains an Abelian normal subgroup with a factor group being isomorphic to the homogeneous orthochronous Lorentz group \cite{Sachsss,newpen65} (see also section 2.5 of Ref. \cite{ANK} for a quick summary). Most crucially, a uniquely defined four-dimensional normal subgroup of the BMS group may be obtained, representing the \emph{four rigid spacetime translations --- thereby providing the basis for a Hamiltonian formalism and the laws for energy-momentum conservation}.

As a quick refresher, the BMS group comprises transformations where the two angular coordinates $\theta$, $\phi$ (which serve as labels for the null generators of the outgoing null hypersurfaces $u=$ constant) undergo a conformal transformation with $K(\theta,\phi)$ being the conformal factor, and $u$ would transform as $u'=K(\theta,\phi)^{-1}(u+\alpha(\theta,\phi))$. The function $\alpha(\theta,\phi)$ is arbitrary, and if expanded using the spherical harmonics basis, then the $l=0,1$ terms would correspond to the Poincar\'{e} translations. If the coordinates $\theta$ and $\phi$ are left unchanged (i.e. no Lorentz rotation occurs), this subgroup of transformations is termed \emph{supertranslations}. If in addition only $l=0,1$ terms are present (so $\alpha(\theta,\phi)$ does not have any spherical harmonic components where $l\geq2$), then we have the subgroup of \emph{translations} having four parameters (one from $l=0$, three from $l=1$). 

\subsubsection{Asymptotically de Sitter spacetimes}\label{Section4B2}

Although the Bondi et al.'s approach has been carried out to include a cosmological constant by He and Cao \cite{chi1}, that effort appeared to have only demonstrated that a non-conformally flat $\mathcal{I}$ would indeed allow for gravitational waves to carry energy away from the source --- culminating in the axisymmetric metric for $\mathcal{I}$ [i.e. Eq. (\ref{scriaxi})], and an expression for $\Psi_4$ having $O(r^{-1})$ terms involving the usual Bondi news --- indicating the presence of outgoing gravitational waves. This work was after Ashtekar et al. studied the conformal structure of asymptotically de Sitter spacetimes and found that the so-called ``usual'' condition of $\mathcal{I}$ being conformally flat would necessarily imply that gravitational waves do not carry energy away from $\mathcal{I}$ \cite{ash1}, and so was perhaps intended to show that the physics of gravitational waves with a cosmological constant is ``meaningful'' after all, when $\mathcal{I}$ is not forced to be conformally flat.

Now, a full treatment \`{a} la Bondi et al. should in principle be able to extend the results to include a cosmological constant, just like how Saw adopted the Newman-Unti approach with $\Lambda\in\R$ \cite{Vee2016,Vee2017}. More specifically, the mass-loss formula in Eq. (\ref{0BondimasslossGraonly}) may well have been produced, if the Bondi approach in Ref. \cite{chi1} went further to work out the supplementary conditions. Moreover, one could also study the permissible coordinate transformations that preserve the form of the asymptotically de Sitter metric. Following that approach, one may expect to find that if $\mathcal{I}$ is conformally flat (which may be characterised by the vanishing of $\sigma^o$ --- see Eqs. (\ref{xi}) above and the discussion that followed), then the asymptotic symmetry group is just the ten-dimensional de Sitter group \cite{ash1}. In the general case where $\mathcal{I}$ is non-conformally flat (i.e. with $\sigma^o\neq0$), there would be a set of coupled PDEs which would determine the allowed transformations (to be presented and discussed elsewhere).


Alternatively, since we have already obtained the metric for $\mathcal{I}$ \cite{Vee2016} (see Section \ref{Section2D} above), we could try to solve the Killing equations for the Killing vectors on $\mathcal{I}$. For simplicity, we could consider the axisymmetric case given by Eq. (\ref{scriaxi}). This would lead to a set of PDEs, which unfortunately, we are unable to analytically solve. Nevertheless, we could look at the axisymmetric topological 2-sphere of constant $u$ given by Eq. (\ref{2axi}), and work with the Killing equations for this 2-surface. We show in Appendix \ref{appendix} that the only Killing vector on this axisymmetric topological 2-sphere is the one associated with axisymmetry \footnote{This is unless $\sigma^o$ satisfies Eq. (\ref{Killingsigma}), which is highly unlikely apart from the trivial $\sigma^o=0$. Even if there is such a non-trivial $\sigma^o$, it would be in a very specialised form --- making it arguably unphysical.}. Therefore, it is plausible to expect no Killing vector for the general topological 2-sphere, and furthermore no Killing vector for the general non-conformally flat $\mathcal{I}$ in Eq. (\ref{metforscri}). This is problematic, as one would not be able to then define conserved quantities that rely on the existence of asymptotic Killing vectors, when an isolated system radiates gravitational waves.

\subsection{ADM mass, conserved quantities}\label{Section4C}

The previous section has shown that with a cosmological constant, there is no asymptotic symmetry when an isolated system radiates gravitational waves, i.e. when $\sigma^o\neq0$. This has the repercussion that any construction of the energy based on asymptotic symmetries would not work. Nevertheless, consider the mass-loss formula which has the general form:
\begin{eqnarray}\label{MP}
\frac{dM}{du}&=&-P,
\end{eqnarray}
where $M(u)$ is the mass of the isolated system and $P(u)$ is the radiation power. If a $u$ (representing time) integral is carried out, then
\begin{eqnarray}
\Delta M=-\int_{u_i}^{u_f}{Pdu},
\end{eqnarray}
i.e. the mass loss of the isolated system over the time interval from $u_i$ to $u_f$ equals the total energy radiated away during that time interval. Written as an indefinite integral, then
\begin{eqnarray}
M=Q-\int{Pdu},
\end{eqnarray}
where $Q$ is a constant with respect to $u$, corresponding to the total mass at some initial time $u_0$. In other words, we have:
\begin{eqnarray}\label{Q}
Q:=M+\int{Pdu}
\end{eqnarray}
defining a conserved quantity $Q$, which remains absolutely conserved as time $u$ proceeds (since it is equal to the initial mass at time $u=u_0$).

Hence from the mass-loss formula in Eq. (\ref{0Bondimasslosspsi0ST}), one may integrate the right-hand side with respect to $u$, and bring it together with the mass $M_\Lambda$ on the left, within the $u$-derivative, to define a conserved quantity. In fact, the Bianchi identity for $D'\Psi_0$ (see Ref. \cite{Vee2016}) gives one such relation of the type given by Eq. (\ref{MP}) for \emph{each order of $r$}. There is also one such relationship from the Bianchi identity for $D'\Psi_1$, other than Eq. (\ref{masslosspreintegration}) that arises from the Bianchi identity for $D'\Psi_2$. Similar remarks apply to the Maxwell equations \cite{Vee2017}.

In the asymptotically flat case, we have the mass-loss formula given by Eq. (\ref{Bondimasslossflat}). Whilst there is a definition of the ADM mass for $\Lambda=0$ \cite{adm,ADMDDD,200years,Poisson} --- corresponding to taking a limit towards spatial infinity $i^0$ instead of null infinity $\mathcal{I}$ ---, we are not aware of any existing work that derives it using the NP formalism since this setup involves null hypersurfaces which naturally go towards $\mathcal{I}$ instead of $i^0$. A suggestion by Saw \cite{Vee2016} conjectures that the conserved quantity $Q$ obtained from Eq. (\ref{Bondimasslossflat}) based on Eq. (\ref{Q}), would perhaps correspond to the ADM mass for asymptotically flat spacetimes. It would be nice to rigorously show this, or to disprove such a proposition. If this turned out to be plausible, then one may regard the corresponding conserved quantity $Q$ obtained from Eq. (\ref{0Bondimasslosspsi0ST}) to be a ``generalised ADM mass that includes $\Lambda$''.

Incidentally, there are 10 conserved quantities for gravitational theory, also known as the 10 NP constants \cite{newpen65a,newpen68,exton}. These are similarly obtained from the Bianchi identities, but are based on the properties of spin-weighted spherical harmonics (of some particular combination of $s$ and $l$) where they vanish when an $\eth$ or $\eth'$ operator acts on them. Such a construction of the 10 NP constants also have a geometrical meaning based on the conformal structure \cite{newpen68}. For the case with $\Lambda\neq0$ however, since the 2-surfaces of constant $u$ on $\mathcal{I}$ are not round spheres, a corresponding description of spin-weighted spherical harmonics on a topological 2-sphere would be needed. Furthermore, it is unclear if an analogous conformal structure with $\Lambda\neq0$ also exists.

%
%

\section{Concluding remarks}\label{Section5}

We have seen how work has progressed over recent years to include a cosmological constant in describing the energy carried by gravitational waves. The structure of null infinity $\mathcal{I}$ gets affected when an isolated system radiates gravitational waves, though it is left unaltered by the emission of electromagnetic radiation. The study to include $\Lambda>0$ by Saw \cite{Vee2016,Vee2017,Vee2017c} has inadvertently also contributed to a description for the $\Lambda<0$ case, where the isolated system determines the structure of the timelike $\mathcal{I}$ such that one cannot impose conformal flatness on $\mathcal{I}$ in order to admit a Bondi news.

There are however, unresolved issues which certainly require further research. In particular, the mass proposals by Saw \cite{Vee2016} as well as by Chru\'{s}ciel and Ifsits \cite{Chrusciel} do not yet have a solid and indisputable physical backing. In fact, one can play around by adding/substracting terms as one wishes on both sides of the equation, in both these proposals. A highly troubling feature that prevents a definitive statement for the expression of the mass with $\Lambda$ has been uncovered and illustrated explicitly in this review article (see Section \ref{Section4B} above), viz. the \emph{lack of asymptotic symmetries}. Without these asymptotic Killing fields, one would be unable to formulate a Hamiltonian framework that depends on such asymptotic symmetries in the exact theory for gravitation. Whilst the linearised theory is able to produce some results using the covariant phase space formalism \cite{ash2,ash3}, so far they are only based on a de Sitter background with the associated Killing fields. Therefore, a worthy consideration to justify or improve the mass proposals from the exact theory would be to carry out a linearised expansion over a non-conformally flat $\mathcal{I}$, viz. Eq. (\ref{metforscri}) [or Eq. (\ref{scriaxi}) for the axisymmetric case]. For instance, by employing the Bondi-Sachs framework as was done by Bishop \cite{gracos1}, one may calculate the energy carried away by gravitational waves from the linearised theory with a \emph{non-conformally flat $\mathcal{I}$}, and hopefully would provide a concrete link with the exact result by Saw, i.e. Eq. (\ref{0Bondimasslosspsi0ST}) above.

Besides that, future work should aim to reconcile the different approaches taken by Saw \cite{Vee2016}, Szabados and Tod \cite{Szabados}, Chru\'{s}ciel and Ifsits \cite{Chrusciel}, as well as by He and Cao \cite{chi1}. The easiest (but non-crucial) task would be using the supplementary conditions to get the mass-loss formula in the Bondi-He-Cao approach and confirm that it is equivalent to Eq. (\ref{masslosspreintegration}) above, as was produced by Saw. Next, it would be fruitful to understand how exactly the Chru\'{s}ciel-Ifsits approach leads to a balance formula for the mass, instead of a mass-loss formula, as well as to relate their renormalised volume term with a similar term appearing in Saw's proposal for the mass in Eq. (\ref{topsph}). Lastly, it would be beneficial to gain an intuitive description of the results by Szabados and Tod arising from their use of twistor techniques especially in relating to physical results, for example the energy carried by gravitational waves.

Since we live in physical spacetime, one could argue that the treatment of gravitational waves could be described purely in terms of the physical spacetime. Ergo, it would be desirable to eventually make clear-cut physical connections, regardless of the approach being undertaken. On top of that, the new conditions on null infinity reported by Xie and Zhang \cite{Zhang} serve as a timely example of a physical spacetime that does not admit a smooth conformal compactification, such that one could possibly conceive some physical spacetimes with different compactification properties.

\appendix

\section{The only Killing vector of an axisymmetric topological $2$-sphere}\label{appendix}

Consider the axisymmetric topological 2-sphere, given by Eq. (\ref{2axi}) (where $u$ is a constant):
\begin{eqnarray}
g=e^{2\Lambda f(\theta)}d\theta^2+e^{-2\Lambda f(\theta)}\sin^2{\theta}d\phi^2.
\end{eqnarray}
Here, we have the spin coefficient $\alpha$ being \cite{Vee2016} \footnote{The use of the spin coefficient $\alpha$ is meant as a shorthand for the expression on the right-hand side. It is not necessary to use $\alpha$ if one prefers to write out those terms explicitly in what follows.}:
\begin{eqnarray}\label{alphaauxi}
\alpha(\theta)=-\frac{1}{2\sqrt{2}\sin{\theta}}\frac{d}{d\theta}\left(e^{-\Lambda f(\theta)}\sin{\theta}\right).
\end{eqnarray}

Let $\vec{X}=X^\theta\vec{\partial}_\theta+X^\phi\vec{\partial}_\phi$ be a Killing vector. Then, the Killing equations are
\begin{eqnarray}
(L_{\vec{X}}g)_{ab}=X^c\partial_cg_{ab}+g_{cb}\partial_aX^c+g_{ca}\partial_bX^c=0,
\end{eqnarray}
giving the three independent equations:
\begin{eqnarray}
\theta\theta&:&\frac{\partial}{\partial\theta}\left(X^\theta e^{\Lambda f}\right)=0\label{Killing1}\\
\phi\phi&:&\frac{\partial X^\phi}{\partial\phi}+X^\theta\cot{\theta}-\Lambda\frac{\partial f}{\partial\theta}X^\theta=0\label{Killing2}\\
\theta\phi&:&\frac{\partial X^\theta}{\partial\phi}+\frac{\partial X^\phi}{\partial\theta}e^{-4\Lambda f}\sin^2{\theta}=0.\label{Killing3}
\end{eqnarray}

The first of these equations give
\begin{eqnarray}\label{Xtheta}
X^\theta(\theta,\phi)=\frac{dA(\phi)}{d\phi}e^{-\Lambda f(\theta)},
\end{eqnarray}
where $A(\phi)$ is an arbitrary function of $\phi$.

The term involving $\Lambda$ in the second of these equations can be eliminated using the first equation, giving (after some simplifications):
\begin{eqnarray}\label{DEXphi}
\frac{\partial X^\phi}{\partial\phi}+\frac{1}{\sin{\theta}}\frac{\partial}{\partial\theta}\left(X^\theta\sin{\theta}\right)=0.
\end{eqnarray}
Substituting $X^\theta$ from Eq. (\ref{Xtheta}) in terms of $A(\phi)$ into Eq. (\ref{DEXphi}), and integrating with respect to $\phi$ would lead to
\begin{eqnarray}
X^\phi(\theta,\phi)=2\sqrt{2}\alpha(\theta)A(\phi)-X(\theta),
\end{eqnarray}
where $X(\theta)$ is an arbitrary function of $\theta$, and we have used $\alpha(\theta)$ from Eq. (\ref{alphaauxi}).

Hence, inserting the general forms of $X^\theta(\theta,\phi)$ and $X^\phi(\theta,\phi)$ into the third equation produces:
\begin{eqnarray}\label{harmonic}
\frac{d^2A(\phi)}{d\phi^2}+\left(2\sqrt{2}e^{-3\Lambda f(\theta)}\frac{d\alpha(\theta)}{d\theta}\sin^2{\theta}\right)A(\phi)=e^{-3\Lambda f(\theta)}\frac{dX(\theta)}{d\theta}\sin^2{\theta}.
\end{eqnarray}
Let $\displaystyle\omega(\theta)^2=2\sqrt{2}e^{-3\Lambda f(\theta)}\frac{d\alpha(\theta)}{d\theta}\sin^2{\theta}$. The homogeneous part of this differential equation (i.e. the left-hand side) is the harmonic oscillator with frequency $\omega(\theta)$, i.e. the complementary function $A(\phi)$ is a linear combination of $\sin{(\omega(\theta)\phi)}$ and $\cos{(\omega(\theta)\phi)}$. But since $A(\phi)$ cannot depend on $\theta$, this solution to the homogeneous part requires $A(\phi)$ to be identically zero. The inhomogeneous term on the right-hand side is a ``constant'' with respect to the variable $\phi$, giving the particular integral $A(\phi)=A$ where $A$ is a constant. However, $A(\phi)$ being a constant implies that $X^\theta=0$ from Eq. (\ref{Xtheta}) and consequently the Killing equations Eqs. (\ref{Killing2}) and (\ref{Killing3}) dictate that $X^\phi=X$, where $X$ is a constant.

Therefore, we have just one single Killing vector on this axisymmetric topological 2-sphere
\begin{eqnarray}
\vec{X}=\vec{\partial}_\phi,
\end{eqnarray}
which is associated with the axisymmetry. In other words, the asymptotic shear $\sigma^o$ destroys the asymptotic symmetry of these 2-surfaces when there is a non-zero cosmological constant $\Lambda$. Well, this is unless $f(\theta)$ satisfies $\omega^2(\theta)=$ constant, i.e.
\begin{eqnarray}\label{Killingsigma}
\displaystyle\omega(\theta)^2=2\sqrt{2}e^{-3\Lambda f(\theta)}\frac{d\alpha(\theta)}{d\theta}\sin^2{\theta}=\textrm{constant},
\end{eqnarray}
where recall that $\alpha(\theta)$ is given by Eq. (\ref{alphaauxi}). But there does not seem to be a solution for $f$ and therefore $\sigma^o$ which would make this possible except for the trivial $f=0$, i.e. $\sigma^o=0$. Moreover, such a specialised form of $\sigma^o(\theta)$ (if a non-trivial solution exists) may not be physically general.

\begin{acknowledgments}
V.-L. Saw is supported by the University of Otago Doctoral Scholarship.
\end{acknowledgments}

\bibliographystyle{spphys}       
\bibliography{Citation}

\end{document}